\documentclass[conference,letterpaper]{IEEEtran} 
\usepackage[utf8]{inputenc}
\usepackage[T1]{fontenc}
\usepackage{microtype}
\usepackage{graphicx}
\usepackage[hidelinks,bookmarks=false]{hyperref}
\usepackage{balance}

\usepackage{booktabs}
\usepackage{array}
\usepackage{colortbl}
\usepackage{multirow}
\usepackage{longtable}

\newcolumntype{v}[1]{>{\raggedright \hspace {0pt}}p{#1}}
\newcolumntype{G}[1]{>{\columncolor{gray90}}#1}

\definecolor{Gray}{gray}{0.8}
\definecolor{gray25}{gray}{0.25}
\definecolor{gray50}{gray}{0.50}
\definecolor{gray75}{gray}{0.75}
\definecolor{gray90}{gray}{0.9}

\newcommand{\grayrow}{\rowcolor{gray90}}
\title{T-Reqs: Tool Support for Managing Requirements in Large-Scale Agile System Development}
\author{
  \IEEEauthorblockN{Eric Knauss\IEEEauthorrefmark{1}, Grischa Liebel\IEEEauthorrefmark{1}, \\ Jennifer Horkoff\IEEEauthorrefmark{1}, Rebekka Wohlrab\IEEEauthorrefmark{1}, Rashidah Kasauli\IEEEauthorrefmark{1}}
  \IEEEauthorblockA{\IEEEauthorrefmark{1}Chalmers $\mid$ University of Gothenburg\\Email: eric.knauss@cse.gu.se}
  \and
  \IEEEauthorblockN{Filip Lange\IEEEauthorrefmark{2}, Pierre Gildert\IEEEauthorrefmark{2}}
  \IEEEauthorblockA{\IEEEauthorrefmark{2}Ericsson AB\\Email: filip.lange@ericsson.se}
}

\begin{document}

\maketitle

\begin{abstract}
T-Reqs is a text-based requirements management solution based on the git version control system.
It combines useful conventions, templates and helper scripts with powerful existing solutions from the git ecosystem and provides a working solution to address some known requirements engineering challenges in large-scale agile system development.
Specifically, it allows agile cross-functional teams to be aware of requirements at system level and enables them to efficiently propose updates to those requirements. 
Based on our experience with T-Reqs, we i) relate known requirements challenges of large-scale agile system development to tool support; ii) list key requirements for tooling in such a context; and iii) propose concrete solutions for challenges.

\end{abstract}


\begin{table*}[t]
\caption{Challenges and requirements for requirements tooling in large-scale agile system development. 
}
\label{tab:chlg-reqt}
\begin{tabular}{v{0.25\textwidth}v{0.15\textwidth}v{0.25\textwidth}v{0.25\textwidth}}
\toprule
\textbf{Challenge} & \textbf{Current status} & \textbf{User story/requirement} & \textbf{Solution in T-Reqs}\tabularnewline
\midrule
\textbf{Updating and deprecating requirements.} Requirements reside between teams on different levels. 
Teams have different scopes but dependencies exist. 
It is difficult to establish governance and policing function for shared requirements. 
& 
Updates of requirements must be proposed to a central role. 
The process is slow, the central role becomes a bottleneck, and changes that appear non-critical may be omitted (\emph{$\rightarrow$C1}).
%
%
& 
\textbf{US1} As a member of a XFT, I want to \\
\textbf{a)} \ldots share new knowledge we learnt about existing requirements during a sprint so that our implementation and the requirements on system level are consistent. \\
\textbf{b)} \ldots be aware of requirements changes that affect my team so that we can pro-actively address dependencies.
& 
Git allows to group changes (e.g. to source code and tests) into commits that then can be pushed to the main branch. 
T-Reqs allows to manage requirements in exactly the same way.
A team pulling the latest changes from the main branch will see conflicts on either of these artefacts as merge conflicts in git.
\tabularnewline
\grayrow \textbf{Access to tooling and requirements.} 
Traditional tools rely on defined change management processes and often do not scale well with respect to parallel users and changes. 
Development teams find this situation at odds with agile practices and pace. 
& 
Often, teams do not have access to tooling and requirements, since licenses are expensive and the number of parallel users  is limited (\emph{$\rightarrow$C1}).
&
\textbf{US2} As a member of a XFT, I want to update system requirements efficiently, without too much overhead, and ideally integrated in the tools I use in my daily work.
&
Development teams are used to git, thus T-Reqs provides them with a familiar interface to manipulate requirements.
{T-Reqs-specific conventions, templates and scripts allow generating specific views and reports for non-technical stakeholders.}
\tabularnewline
\textbf{Consistent requirements quality.} Quality of requirements 
(i.e. user stories, backlogs) differs (e.g. in detail). 
This allows best requirements practices for each domain, but reasoning on system level is difficult \cite{Wohlrab2018}.
&
No appropriate review and alignment process exists that would allow to include an individual team's way of working (\emph{$\rightarrow$C1-2}).
&
\textbf{US3} As a system manager, I want to make sure that proposed updates to requirements are of good quality, do not conflict with each other, or with the product mission.
&
Many organisations that rely on git are also using gerrit to manage reviews (e.g. of source code). 
T-Reqs organizes requirements in a way that allows to do that with requirements as well.
\tabularnewline
\grayrow
\textbf{Managing experimental requirements.} When exploring new functionality or product ideas, experimental requirements need to be treated differently from stable requirements. 
 Still, they need to be captured and later integrated in the system view \cite{Kasauli2017a}.
 & 
 The requirements database is cloned before experimenting and must be blocked for other changes when the clone is eventually be ported back ($\rightarrow$\emph{C2}). 
 & 
\textbf{US4} As member of an experimenting team, I want to experiment with new requirements and features so that I can better assess their business value and cost. 
This must not affect existing requirements during the experiment or block the requirements database afterwards.
& 
Git allows creating branches to experiment with requirements, but also with models and source code. 
Git merge and gerrit help to merge branches, without blocking the main database.
Merge conflicts will directly relate to requirements conflicts, since requirements are stored line-wise.
\tabularnewline
\textbf{Create and maintain traces.} 
In continuous integration and delivery, agile teams struggle to provide sufficient tracing to allow determining the status of individual features. 
&
Tracing 
does not offer direct value to agile teams and is  
not integrated in their workflows (\emph{$\rightarrow$C3}). 
&
\textbf{US5} As a member of a XFT, I want to maintain traces between requirements, change sets, and tests in a way that is integrated with my natural workflow and enables valuable feedback.
&
Git automatically links changes of code and requirements in commits. 
{T-Reqs adds conventions, templates, and scripts for additional finer-grained tracing.}
Providing cross-functional teams with feedback based on tracing information can further motivate good trace link quality.
\tabularnewline
\grayrow \textbf{Plan verification and validation based on requirements.} 
When requirements updates by individual teams are not shared on system level, it is impossible to plan verification and validation pro-actively. 
& 
Requirements changes are difficult to share (\emph{$\rightarrow$C1-3}) and the need to update complex system testing infrastructure may surface late.
&
\textbf{US6} As a test architect or system manager, I want to be aware of new requirements for the test infrastructure early on so that I can plan verification and validation pro-actively. 
&
T-Reqs suggests a suitable review process via gerrit that has proven to spread information about critical changes effectively between key stakeholders.
Ease of use makes it more likely that teams share requirements updates in a timely manner. 
\tabularnewline
\bottomrule
\end{tabular}
\end{table*}

\section{Introduction}
Requirements engineering (RE) is crucial to support agile development of large systems with long lifetimes. 
Yet, traditional tooling does not sufficiently support addressing known RE challenges in large-scale agile \cite{heikkila2015mapping,heikkila2017managing,inayat2015systematic}.
In particular, it is hard to cross the boundaries between three domains in large-scale agile organizations: the customer facing domain, the development domain, and the system domain \cite{Kasauli2017a}.
Agile cross-functional teams must be aware of requirements at system level and able to efficiently propose updates to those requirements.
%
%
%
%
%
Based on real world experience with selecting, rolling out, and, for several releases now, using a novel tool solution based on git and markdown, we  i) relate known requirements challenges of large-scale agile system development to tool support; ii) list key requirements for tooling in such a context; and iii) propose concrete solutions for challenges.
%

\section{Industrial Context}

We report here based on the experience from one particular product developed at a specific department within Ericsson. 
The product development organization of the department has changed significantly in recent years. 
Before 2012, the development process defined two releases per year, requirements were created upfront, and were defined in IBM Rational Requisite Pro (first released in 1995). 
In 2012, the department started a transformation towards agile methods and continuous integration at scale. 
In 2017, this transformation has reached a state where the product is released once per month.

While it is hard to quantify the level of agility in the company, the changes are significant and comprehensive. 
The development process is based on a continuous feature flow and allows about 30 small, cross-functional development teams (XFT) to develop software following a Scrum approach.
This has significantly changed the way of managing requirements.
%
%
The following three changes stick out:
%
%

\noindent
\begin{center}
\footnotesize
\begin{tabular}{|lv{0.86\columnwidth}|}
\hline
C1 & \emph{Requirements are updated more frequently.} This can cause teams to block each other.\tabularnewline
C2 & \emph{Number of database clones grows.} The need to clean up databases because of unfinished requirements as well as to frequently port requirements back to a main database causes significant effort.\tabularnewline
C3 & \emph{No direct link between requirements and commits.} 
It is difficult to determine when a feature has been delivered.\tabularnewline
\hline
\end{tabular}
\end{center}

The existing requirements tool was no longer deemed sufficient and the department evaluated 16 different requirements engineering tools.
None was found to satisfy their specific needs. 
Two requirements stuck out as especially hard to fulfill:
\noindent
\begin{center}
\footnotesize
\begin{tabular}{|lv{0.86\columnwidth}|}
\hline
R1 & 
\emph{The tool must use git as version control system or must support easy synchronisation with git.} Teams are working with git and since it is their responsibility to propagate changes to requirements as part of a sprint, a suitable requirements tool must allow so within the development context.\tabularnewline
R2 & \emph{The tool must support simultaneous work of many users on the same artefact.} Since several teams may be working in the scope of particular requirements, they must be able to report their changes without reserving the artefact and blocking each other.\tabularnewline
\hline
\end{tabular}
\end{center}

When searching for a suitable tool, 
an in-house proposal proved to provide the best fit: T-Reqs.
%
%
%
T-Reqs (textual requirements) suggests managing requirements in markdown format in text files within git\footnote{{Demonstrator available at \url{https://github.com/regot-chalmers/treqs}}}.
%
%
%
As depicted in Table \ref{tab:chlg-reqt}, this surprisingly simple solution satisfies the majority of concerns and fulfils the critical requirements {with only small additions (scripts and templates) beyond existing solutions in the git ecosystem}. 
{These allow for example to generate reports and views on the requirements and related artifacts with tracing links.}

\section{T-Reqs vs. Challenges and Requirements}
Recently, a growing number of empirical papers discusses requirements engineering challenges in agile development \cite{inayat2015systematic,heikkila2015mapping,Kasauli2017a} as well as the need to identify new ways of managing requirements in agile organizations \cite{heikkila2017managing}.
In Table \ref{tab:chlg-reqt}, we select  a subset of these challenges that is especially relevant to tooling in our company context.
We also extract requirements towards tooling in form of user stories and discuss how T-Reqs helps to address these challenges.



\section{Discussion and Outlook}

In this paper\footnote{{Partially supported by Software Center: \url{https://www.software-center.se}}}, we present T-Reqs, an approach to rely on textual requirements based on a markdown format and to manage those in git. 
We demonstrate that this approach addresses critical challenges in large-scale agile system development.
%
%
T-Reqs is proven in practice, especially in an environment that can rely mainly on textual requirements. 
However, models do exist and those cannot be as easily managed with the proposed solution, asking for future research.
Given the specific environment from which we draw our experience, we cannot reason about the value of T-Reqs in other industrial contexts. 
Regardless, we believe that we can facilitate a discussion of changing needs towards tooling in agile development.


\balance

\bibliographystyle{IEEEtranS}
\bibliography{2018-RE-TReqs}

\begin{thebibliography}{1}
\providecommand{\url}[1]{#1}
\csname url@samestyle\endcsname
\providecommand{\newblock}{\relax}
\providecommand{\bibinfo}[2]{#2}
\providecommand{\BIBentrySTDinterwordspacing}{\spaceskip=0pt\relax}
\providecommand{\BIBentryALTinterwordstretchfactor}{4}
\providecommand{\BIBentryALTinterwordspacing}{\spaceskip=\fontdimen2\font plus
\BIBentryALTinterwordstretchfactor\fontdimen3\font minus
  \fontdimen4\font\relax}
\providecommand{\BIBforeignlanguage}[2]{{%
\expandafter\ifx\csname l@#1\endcsname\relax
\typeout{** WARNING: IEEEtranS.bst: No hyphenation pattern has been}%
\typeout{** loaded for the language `#1'. Using the pattern for}%
\typeout{** the default language instead.}%
\else
\language=\csname l@#1\endcsname
\fi
#2}}
\providecommand{\BIBdecl}{\relax}
\BIBdecl

\bibitem{heikkila2015mapping}
V.~T. Heikkil{\"a}, D.~Damian, C.~Lassenius, and M.~Paasivaara, ``A mapping
  study on requirements engineering in agile software development,'' in
  \emph{41st Conf. on Softw. Eng. and Adv. Appl. (SEAA)}, 2015, pp. 199--207.

\bibitem{heikkila2017managing}
V.~T. Heikkil{\"a}, M.~Paasivaara, C.~Lasssenius, D.~Damian, and C.~Engblom,
  ``Managing the requirements flow from strategy to release in large-scale
  agile development: a case study at ericsson,'' \emph{Empirical Software
  Engineering}, pp. 1--45, 2017.

\bibitem{inayat2015systematic}
I.~Inayat, S.~S. Salim, S.~Marczak, M.~Daneva, and S.~Shamshirband, ``A
  systematic literature review on agile requirements engineering practices and
  challenges,'' \emph{Computers in Human Behavior}, vol.~51, pp. 915--929,
  2015.

\bibitem{Kasauli2017a}
R.~Kasauli, G.~Liebel, E.~Knauss, S.~Gopakumar, and B.~Kanagwa, ``Requirements
  engineering challenges in large-scale agile system development,'' in
  \emph{Proc. of 25th Reqts. Eng. Conf. (RE)}, Lisbon, Portugal, 2017.

\bibitem{Wohlrab2018}
R.~Wohlrab, P.~Pelliccione, E.~Knauss, and S.~Gregory, ``The problem of
  consolidating {RE} practices at scale: An ethnographic study,'' in
  \emph{Proc. of 24th Int. Working Conf. on Reqts. Eng.: Foundation for
  Software Quality (REFSQ)}, Utrecht, The Netherlands, 2018.

\end{thebibliography}

\end{document}